\documentclass[aps,prl,floatfix,twocolumn]{revtex4-1}
\usepackage{amsmath,amssymb}
\usepackage{graphicx}
\usepackage{epsfig}
\usepackage[usenames]{color}

\usepackage{ulem}
\usepackage{xcolor}

\begin{document}

\title{Quantum Anomalous Valley Hall Effect for Bosons}



\author{V.~M.~Kovalev and I.~G.~Savenko}
\affiliation{A.V.~Rzhanov Institute of Semiconductor Physics, Siberian Branch of Russian Academy of Sciences, Novosibirsk 630090, Russia\\
Center for Theoretical Physics of Complex Systems, Institute for Basic Science (IBS), Daejeon 34126, Korea}


\date{\today}

\begin{abstract}
We predict the emergence and quantization of the valley Hall current of indirect excitons residing in noncentrosymmetric two-dimensional materials in the absence of an external magnetic field. 
Thus we report on the quantum anomalous valley Hall effect (QAVHE) for bosons.
The crucial ingredients of its existence are the finite anomalous group velocity of exciton motion 
and the presence of the Bose-Einstein condensate, from which the particles are pushed out by external illumination to contribute to the stationary nonequilibrium  current, which exhibits a quantized behavior as a function of the frequency of light.
\end{abstract}


\maketitle

The essence of the ordinary Hall effect, which discovery is tracing back to 1879, is in the emergence of particle transport in the transverse direction under an external magnetic field~\cite{RefHall}. 
If the magnetic field becomes sufficiently strong, the Hall conductance takes on the quantized values~\cite{RefAndo, RefKliz}; such is the nature of the quantum Hall effect (QHE).
In general, all the Hall effects are based on the time reversal symmetry breaking employing either an external field, or internal magnetization, or spin-orbital coupling. 
For instance, there takes place the anomalous Hall effect (AHE) in certain materials 
due to the presence of magnetic polarization and spin-orbit splitting~\cite{RefKar}. 
The AHE also has its quantized counterpart~\cite{RefAnS}.

%

Anomalous transport of carriers of charge in low-dimensional structures is a  captivating fundamental and a simultaneously application-oriented subject of research, which is recently attracting growing attention~\cite{RefNagaosa}.
One of the conceptual problems here is the study of the AHE resulting from the topological properties of the carriers of charge~\cite{RefHasan} due to the emergence of the Berry phase associated with closed trajectories in the parameter space~\cite{RefBerry}.
It turns out that the Berry curvature plays a central role in various transport phenomena in condensed matter physics~\cite{RefChang, RefSundaram, BerryRMP}, one of the most intriguing of which is the valley Hall effect (VHE)~\cite{RefThouless, RefVHE1, RefVHE2, RefVHE3}.

In intuitively apprehensible quasiclassical language, the origin of the anomalous Hall transport lies in a specific term in the expression of the particle group velocity, $\dot{\textbf{r}}=\partial_\textbf{p}\varepsilon(\textbf{p})-\dot{\textbf{p}}\times {\bf \Omega}_\textbf{p}$,
where $\varepsilon(\textbf{p})$ is the  dispersion relation and ${\bf \Omega}_\textbf{p}$ is the Berry curvature, expressed via the periodic amplitude of the Bloch wave function $|u\rangle$: ${\bf \Omega}_\textbf{p}=\partial_\textbf{p}\times\langle u|i\partial_\textbf{p}|u\rangle$.
For a charged particle subject to an electric field $\textbf{E}(t)$, the quasiclassical equation of motion reads $\dot{\textbf{p}}=e\textbf{E}(t)$, giving the anomalous velocity term $-e\textbf{E}(t)\times{\bf \Omega}_\textbf{p}$ acting as a pseudomagnetic field in momentum space. This very term is responsible for the anomalous Hall currents.

We pose a question: Is there any influence of the Berry curvature on dynamics of overall neutral compound particles like bosonic excitations in semiconductor crystals, called excitons (electron-hole pairs)?
For the first glance, it seems to vanish due to the insensitivity of their center-of-mass motion to external uniform electromagnetic fields.
In the meantime, transport of neutral particles is of great interest, and also it would be highly beneficial to map the topological properties of fermions on bosons.
Excitons not only play an important role in optical phenomena occurring in low-dimensional structures~\cite{RefButov2, RefButov3} but also they can be widely used in applications~\cite{RefButov1}.
One of the advantages of excitonic devices is the possibility of exciton-photon coupling, which is essential for the manipulation of optical signals, including the opportunity to form coherent quasi-condensate states with the vanishing sensitivity to a disorder present in any nanostructure.

In this Letter, we address this question and show that in the presence of the two-dimensional Bose-Einstein quasi-condensate (later condensate) and finite Berry phase, the exciton current density becomes quantized in the absence of an external magnetic field; we call this phenomenon the \textit{Quantum Anomalous Valley Hall Effect} (QAVHE) for bosons. 
It manifests itself in exciton current steplike behavior as a function of frequency of external polarized light, which breaks the time reversal symmetry.
%

Excitons can form from electrons and holes sharing a common quantum well (direct excitons) or separated from each other, thus residing in a double quantum well (indirect excitons).
The main feature of indirect excitons is that they possess a built-in dipole moment.
The most well-studied system of materials for double quantum wells and thus for the study of optical and transport properties of excitons is GaAs and its alloys with aluminum and indium.
The discovery of novel purely 2D materials, such as monomolecular layers of transition metal dichalcogenides (TMDs), benchmarked new chapter in both the direct~\cite{RefGlazov, RefYou, RefRadisavljevic, RefSundaram2, xu2014spin} and indirect~\cite{ButovTMDs1, ButovTMDs2, ButovTMDs3, ButovTMDs4} exciton physics.
These monolayers have a complex  Brillouin zone~\cite{RefVHE1}, containing two inequivalent valleys located at the opposite edges.
Due to sizeable intervalley distance in momentum space, the quantum number indicating the valley is robust against external perturbations, thus creating an additional degree of freedom similar to spin or charge of the particles.
This paradigm forms the basis of bosonic valleytronics.

Another property of indirect excitons, which distinguishes them from their direct counterparts, is a long lifetime, typically ranging from nanoseconds to microseconds.
The lifetime is long due to the weak overlap of spatially separated electron and hole wave functions.
Fast energy relaxation allows for cooling indirect excitons down to $\sim 0.1$~K within their lifetime, that is well below the temperature of quantum degeneracy of exciton gas, thus opening an experimental way to form a coherent exciton state -- the excitonic quasi-condensate.
Experiments shows typical temperatures of exciton condensation $1-5$~K in GaAs nanostructures.
Recently there have been reported reasonable arguments to expect a much higher condensation temperature for indirect excitons in TMD monolayers such as MoS$_2$~\cite{RefButov2}, which makes these materials very promising for the observation of quantum coherent phenomena at higher temperatures.

\begin{figure}[!t]
\includegraphics[width=0.49\textwidth]{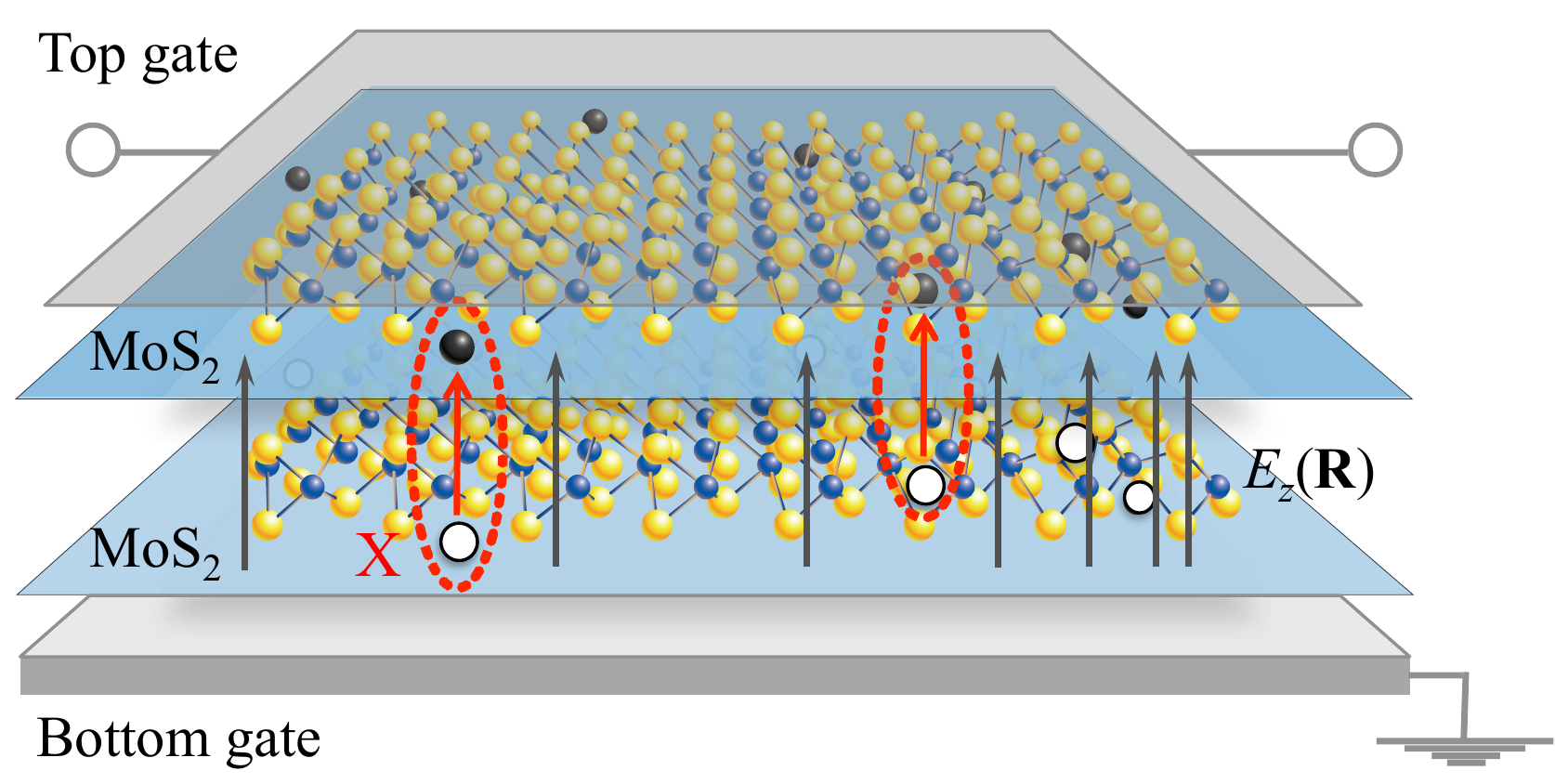}
\caption{System schematic. 
A two-dimensional exciton gas localized in two MoS$_2$ layers and exposed to an electrostatic field $E_z(\textbf{R})$ by means of the top an bottom gates. 
The field is inhomogeneous in the in-plane directions, which is shown by black arrows becoming more dense from left to the right.}
\label{Fig1}
\end{figure}

If we look at the quasiclassical picture, the dipole moment of an exciton $p=ed\,(e>0)$, where $d$ is the interlayer separation, directs across the layers (see Fig.~\ref{Fig1}). 
If a voltage is applied across the layers, it generates a static electric field $E_z$, resulting in the exciton energy shift by $U=-pE_z$ and thus giving an opportunity to create a force $\textbf{F}(\textbf{R})=-\nabla_{\textbf{R}}U(\textbf{R})$ which influences the center-of-mass motion due to the change of the electric field along the layers $E_z(\textbf{R})$ \cite{RefButov1}.
The presence of the Berry phase results in an anomalous contribution to the exciton center of mass velocity \cite{ExcitonBerry1, ExcitonBerry2, ExcitonBerry3}, proportional to $-[\nabla_{\textbf{R}}U\times{\bf \Omega}_{ex}]$, where ${\bf \Omega}_{ex}$ is the exciton Berry curvature.
This contribution is responsible for the exciton QAVHE.


The Heisenberg equations of motion,
\begin{gather}\label{Equation1}
\dot{\textbf{R}}=\partial_{\textbf{P}}H,\,\,~~~\dot{\textbf{P}}=-\partial_{\textbf{R}}H,
\end{gather}
govern the transport of an individual exciton, where $\textbf{R}$ and $\textbf{P}$ are the exciton center-of-mass coordinate and momentum.
Each exciton in a TMD consists of two Coulomb-interacting Dirac particles.

\begin{figure}[!b]
\includegraphics[width=0.49\textwidth]{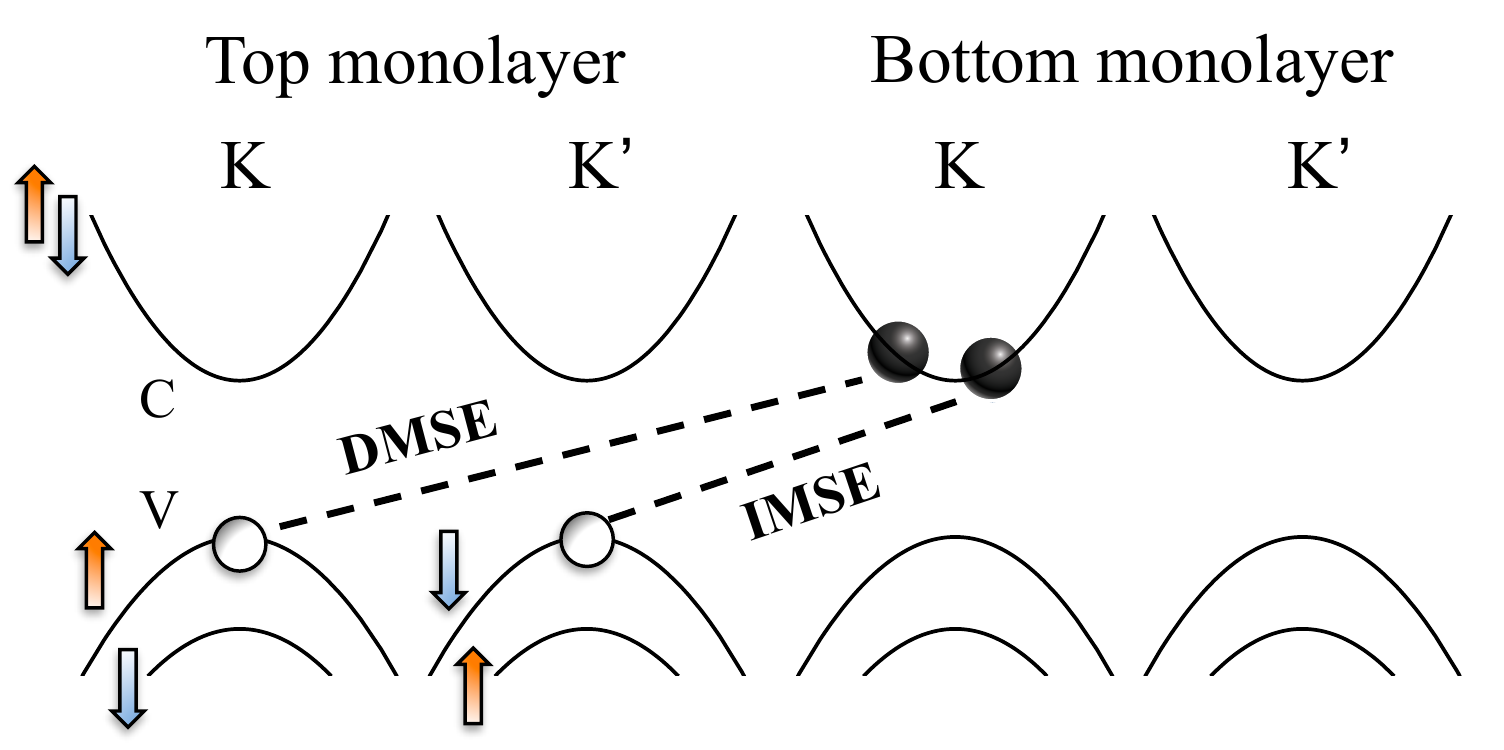}
\caption{Two types of indirect excitons in momentum space: the definition of the direct momentum space exciton (DMSE) and indirect momentum space exciton (IMSE).}
\label{Fig2}
\end{figure}

The two-valley structure of MoS$_2$ Brillouin zone permits the existence of two types of indirect excitons.
If an electron from the top layer and a hole of the bottom layer inhabit the same valley, such an exciton is considered to be direct in momentum space (we will refer to it as DMSE).
The exciton is indirect in momentum space (we will use the abbreviation IMSE) if the particles reside in different valleys (Fig.~\ref{Fig2}).
Below we consider the general case and later discuss how these two types of excitons participate in the QAVHE. 

The Hamiltonian describing an exciton in the TMD layers in the continuous model reads~\cite{ExcitonBerry2, BerryTransformation1}
\begin{gather}\label{Equation2}
H=H_0(\hat{\textbf{p}}_e)+H_0(\hat{\textbf{p}}_h)+u(\textbf{r}_h-\textbf{r}_e)+U(\textbf{r}_e,\textbf{r}_h).
\end{gather}
%
In the vicinity of the conduction and valence band edges, both the $H_0$ terms can be approximated by the parabolic term $H_0(\textbf{p})=\textbf{p}^2/2m$.
In the framework of the two-band Dirac model, which describes well the low-energy spectrum of a TMD monolayer, the electron and hole have equivalent effective masses.
The account of the higher energy bands results in a difference of the particle masses.
However, for our purpose, which is the proof of principle, the two-band model is sufficient.

The second term in  Hamiltonian~\eqref{Equation2} describes the Coulomb interaction between the electron and hole, and $U(\textbf{r}_e,\textbf{r}_h)$ term corresponds to the exciton energy shift due to the voltage applied across the layers.
The Berry-phase effect is introduced via the gauge transformation
%
\begin{gather}\label{Equation3}
\textbf{r}_\alpha\rightarrow\textbf{r}_\alpha+\frac{1}{2}{\bf \Omega}_\alpha(\textbf{p}_\alpha)\times \textbf{p}_\alpha,
\end{gather}
where $\alpha=e,h$. 
%
%
%
%
%
%
%
%
%
%
%
%
%
%
%
Substituting~(\ref{Equation3}) in~(\ref{Equation2}) and then in~(\ref{Equation1}), one finds the equation of (quasiclassical) exciton motion (see the Supplemental Material~\cite{SM} for the details of calculation),
\begin{gather}\label{Equation4}
\dot{\textbf{R}}=\textbf{P}/M-\frac{1}{8}[\textbf{F}_0\times{\bf \Omega}_{ex}],
\end{gather}
where $\textbf{F}_0=\textbf{F}(0)$ is the in-plane static force acting on the exciton center-of-mass and ${\bf \Omega}_{ex}=2{\bf \Omega}_{e}$ is the exciton Berry curvature, taken at the band edges.
The corresponding exciton current reads
\begin{gather}\label{Equation5}
\textbf{j}=\sum_{\textbf{P}}\dot{\textbf{R}}f(\textbf{P},\textbf{R},t).
\end{gather}
In equilibrium, the distribution function is an even function of the center-of-mass momentum $f_0(-\textbf{P})=f_0(\textbf{P})$.
Therefore only the second term in~(\ref{Equation4}) if substituted in (\ref{Equation5}) gives a non-vanishing contribution.
If condensed, the excitons occupy the lowest energy state with zero momentum, $f_0(\textbf{P})=n_c\delta_{\textbf{P},0}$, where $n_c$ is a condensate density. The resulting equilibrium valley Hall current reads
\begin{gather}\label{Equation6}
\textbf{j}_{eq}=-\frac{n_c}{8}[\textbf{F}_0\times{\bf \Omega}_{ex}].
\end{gather}
%


%
%
%
\begin{figure}[!t]
\includegraphics[width=0.49\textwidth]{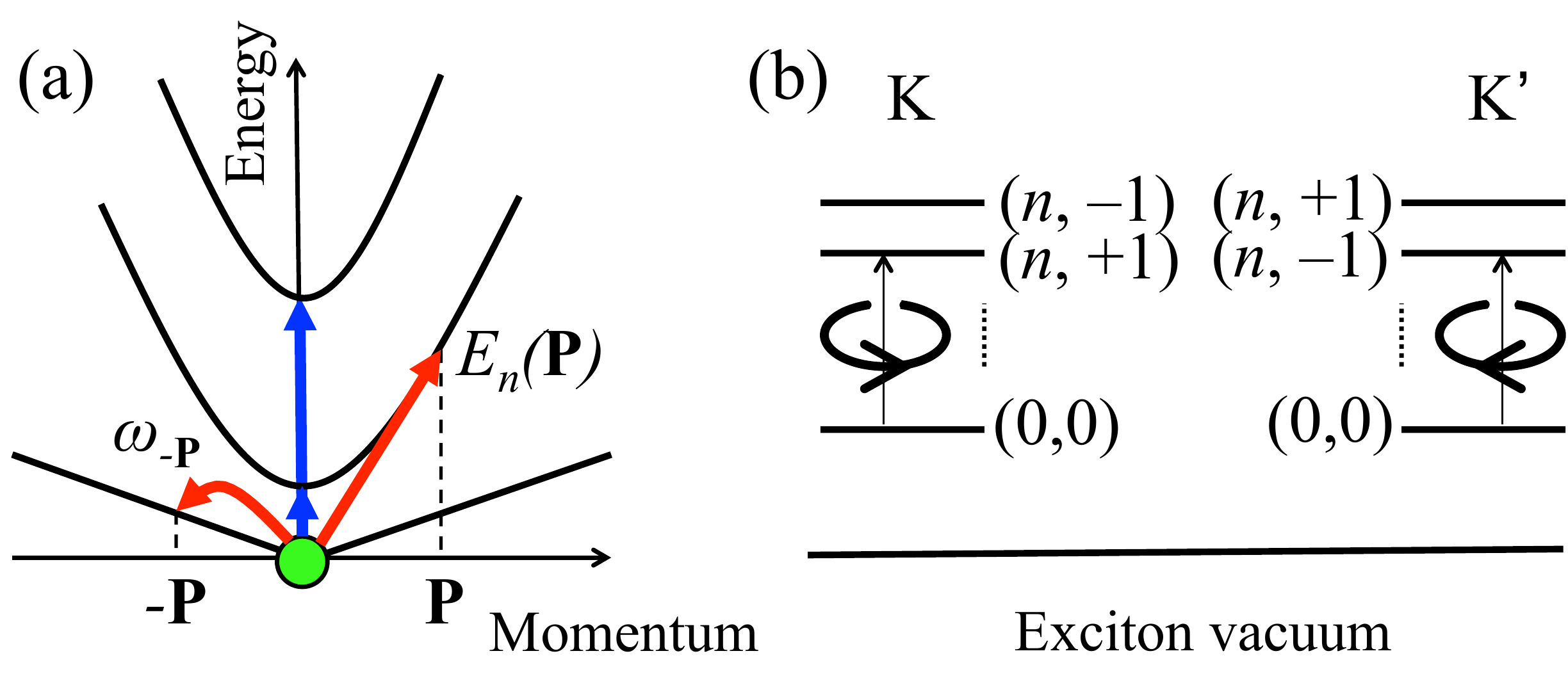}
\caption{
(a) Two principal types of photoinduced transitions in the system. Blue arrows show the direct transitions from the quasi-condensate to excited states, whereas indirect transitions accompanied by the emissions of Bogoliubov quasiparticles with frequencies $\omega_{-\mathbf{P}}$ are depicted by red arrows. 
These processes result in the emergence of two components of the electric current density presented in Fig.~\ref{Fig4} (blue and red curves).
(b) Photoinduced intra-exciton transitions in the system under the circular-polarized light.
}
\label{Fig3}
\end{figure}
%
%
%


Under the action of an external electromagnetic field the exciton distribution acquires the nonequilibrium correction, $f(\textbf{P},\textbf{R},t)=f_0(\textbf{P})+\delta f(\textbf{P},\textbf{R},t)$, describing the population of noncondensed excitonic states $|n,\textbf{P}\rangle$ with $n$ being the quantum number of the internal exciton quantized energy levels and $\textbf{P}\neq0$.
In the stationary regime, the current of photoexcited excitons is determined by the time-averaged distribution function $\delta f_n(\textbf{P})=\overline{\delta f(\textbf{P},\textbf{R},t)}$, yielding
\begin{gather}\label{Equation7}
\textbf{j}=-\frac{1}{8}[\textbf{F}_0\times{\bf \Omega}_{ex}]\sum_{n,\textbf{P}}\delta f_n(\textbf{P}),
\end{gather}
where $\delta f_n(\textbf{P})=\tau R$, with $\tau$ the exciton relaxation time in the excited state,
\begin{eqnarray}\label{Equation8}
R&=&\frac{2\pi}{\hbar S}|\langle n, \textbf{P}|\hat{W}|BEC\rangle|^2
\\
\nonumber
&&\times
\delta(E_{BEC}(n,\textbf{P})-E_{BEC}-\omega)
\end{eqnarray}
is the rate of exciton transitions to the final state  $|n,\textbf{P}\rangle$ from the equilibrium Bose-condensed state $|BEC\rangle$ (per second per unit area).
Here $\omega$ is the frequency of external electromagnetic field, $E_{BEC}$ is the energy of exciton system in BEC regime and $E_{BEC}(n,\textbf{P})$ is the energy of the excited state consisting of the condensate and a photoexcited exciton.
We use a constant $\tau$ assuming the scattering of excited excitons via the short-range impurities.
The Hamiltonian describing the exciton interaction with the external uniform electromagnetic field $\textbf{E}$ is taken in the dipole approximation, $W(\textbf{r})=-e(\textbf{r}\textbf{E})$, where $\textbf{r}=\textbf{r}_h-\textbf{r}_e$ is the relative in-plane electron-hole coordinate.

The exciton energy is characterized by the momentum $\textbf{P}$ and discrete energy levels $\epsilon_n$ of internal exciton motion, $E_n(\textbf{P})=\textbf{P}^2/2M+\epsilon_n$.
We assume that the condensate occurs at the energy state with zero center-of-mass momentum $\textbf{P}=0$ and at the lowest energy level of the internal exciton motion $\epsilon_{n=0}=0$.
According to the Bogoliubov model of a weakly-interacting Bose gas, the low-energy properties of a Bose condensate can be characterized by the excitations having the dispersion $\omega_\textbf{k}=sk\sqrt{1+(k\xi)^2}$, where $s=\sqrt{gn_c/M}$ is a phase velocity, $\xi=1/2Ms$ is a healing length, and $g$ is the strength of exciton-exciton interaction.
Following the Bogoliubov theory, the Bose-condensed state reads
\begin{eqnarray}
\label{Equation9}
\Psi_{BEC}&=&\frac{a_0}{\sqrt{S}}\psi_0(\textbf{r})\\
\nonumber
&&+\psi_0(\textbf{r})\sum_{\textbf{P}\neq 0}\left(u_\textbf{P}a_\textbf{P}+v_{\textbf{P}}a^\dagger_{-\textbf{P}}\right)\frac{\exp(i\textbf{PR})}{\sqrt{S}}.
\end{eqnarray}
Here $a_0=\sqrt{n_cS}$, the first term describes the macroscopically-occupied BEC state, whereas the second term corresponds to the fluctuations with the dispersion $\omega_\textbf{P}$, and $u_\textbf{P},v_{\textbf{P}}$ are the Bogoliubov transformation coefficients.
The function $\psi_n(\textbf{r})$ is the eigen function of internal exciton motion, while the excited states with energies $E_n(\textbf{P})$ have the eigen functions $\Psi_n(\textbf{R},\textbf{r})=\psi_n(\textbf{r})\exp(i\textbf{PR})/\sqrt{S}$.

Due to the presence of two terms in Eq.~(\ref{Equation9}), the matrix element of external field $W(\textbf{r})$ also consists of two terms describing two types of processes, schematically shown in Fig.~\ref{Fig3}(a).
The matrix element of the first-type processes reads~\cite{SM}
\begin{eqnarray}\label{Equation10}
M_n(\textbf{P})&=&a_0W_{n,0}\frac{(2\pi\hbar)^2}{S}\delta(\textbf{P}),~\textrm{where}\\
\nonumber
&&W_{n,0}=\int d\textbf{r}\psi^\dagger_n(\textbf{r})W(\textbf{r})\psi_0(\textbf{r}).
\end{eqnarray}
It corresponds to direct exciton transitions from the condensate to a non-condensed state with the excitation of an internal degree of freedom of individual exciton.
The corresponding photoexcited Hall current reads
\begin{gather}\label{Equation11}
\textbf{j}^{(1)}=[{\bf \Omega}_{ex}\times \textbf{F}_0]\left(\frac{n_c\tau^2}{4\hbar^2}\right)\sum_{n\neq0}
\frac{|W_{n,0}|^2}{1+(\hbar\omega-\epsilon_n)^2\tau^2/\hbar^2}.
\end{gather}
It has a resonant structure and is proportional to the condensate density.
%
%
%
%
%
%

The processes of the second type are described by the second term in Eq.~(\ref{Equation9}), reflecting the  density fluctuations of the BEC (the Bogoliubov quanta).
At zero (small) temperature this excitation branch is (nearly) empty thus the only term containing the creation operator $a^\dagger_{-\textbf{p}}$ contributes to the optical transitions.
This corresponds to the exciton excitation by the photon absorption accompanied by the emission of a Bogoliubov quasiparticle, as indicated in Fig.~\ref{Fig3}(a).
Contrary to the processes of the first type, here the exciton transfers into the non-condensate state with a \textit{nonzero} momentum $\textbf{P}\neq0$ from the condensate state with $\textbf{P}=0$.
Such indirect transitions may occur only if the momentum $\textbf{P}$ is carried away by a ``third body''. The Bose condensate here plays this very role, carrying away the momentum by means of the Bogoliubov excitations with the energy $\omega_{-\textbf{P}}$.

The matrix element describing these processes reads
\begin{gather}\label{Equation12}
M_n(\textbf{P}',\textbf{P})=W_{n0}\frac{(2\pi\hbar)^2}{S}v_{\textbf{P}}\delta(\textbf{P}'-\textbf{P}).
\end{gather}
In the most interesting limit of linear dispersion of Bogoliubov excitations ($\omega_\textbf{P}\approx sP\gg P^2/2M$), we can carry out analytical calculations~\cite{SM} and find the
corresponding contribution to the exciton Hall current:
\begin{eqnarray}\label{EquationCurrent}
\textbf{j}^{(2)}&=&[{\bf \Omega}_{ex}\times \textbf{F}_0]\left(\frac{M\tau}{16\pi\hbar^3}\right)\\
\nonumber
&&\times\sum_{n\neq0}
|W_{n,0}|^2\left(\frac{\pi}{2}+\arctan \left[\frac{(\hbar\omega-\epsilon_n)\tau}{\hbar}\right]\right).
\end{eqnarray}
Evidently, this term vanishes if we disregard the Berry phase effect (the term $\mathbf{j}^{(1)}$ is also zero in this case). 
Note, that it is the arctangent function which ultimately determines the behavior of this component of the electric current. 
Another important property of~\eqref{EquationCurrent} is that it does not explicitly depend on the condensate density, in contrast with~\eqref{Equation11}, even though $\mathbf{j}^{(2)}$ also vanishes in the absence of the condensate.  
The system response~\eqref{EquationCurrent} is a unique characteristic of bosonic systems, making the QAVHE here conceptually different from the conventional fermionic VHE. 
Thus, we conclude that condensation is the necessary condition for the current quantization in the absence of an external magnetic field.
%
%
%
%
%
%


In an experiment, one creates the exciton gas when the sample is illuminated by  light with the frequency exceeding the material band gap.
Due to the fast energy relaxation, the photo-excited electrons and holes cool down and form excitons in the monolayers.
We can expect that both the valleys will be populated with DMSE and IMSE nearly equally.
The Berry curvature of IMSE ${\bf \Omega}_{ex}=0$ since the signs of the Berry curvatures of electrons and holes located in different valleys are opposite.
Thus, only DMSE will contribute to the effect.
Due to the nearly equivalent population of the valleys, the equilibrium valley Hall current vanishes since ${\bf \Omega}^K_{ex}=-{\bf \Omega}^{K'}_{ex}$. However, the photoinduced part of the valley Hall current might be nonzero.
Due to the axial symmetry of the Coulomb electron-hole interaction potential $u(\textbf{r})$, the eigen states of the internal exciton motion are characterized by the principal (radial) quantum number and the projection of the angular momentum: $n=(n_r,m)$.
The axial symmetry dictates the degeneracy of quantum states $(n_r,\pm m)$.
\begin{figure}[!t]
\includegraphics[width=0.49\textwidth]{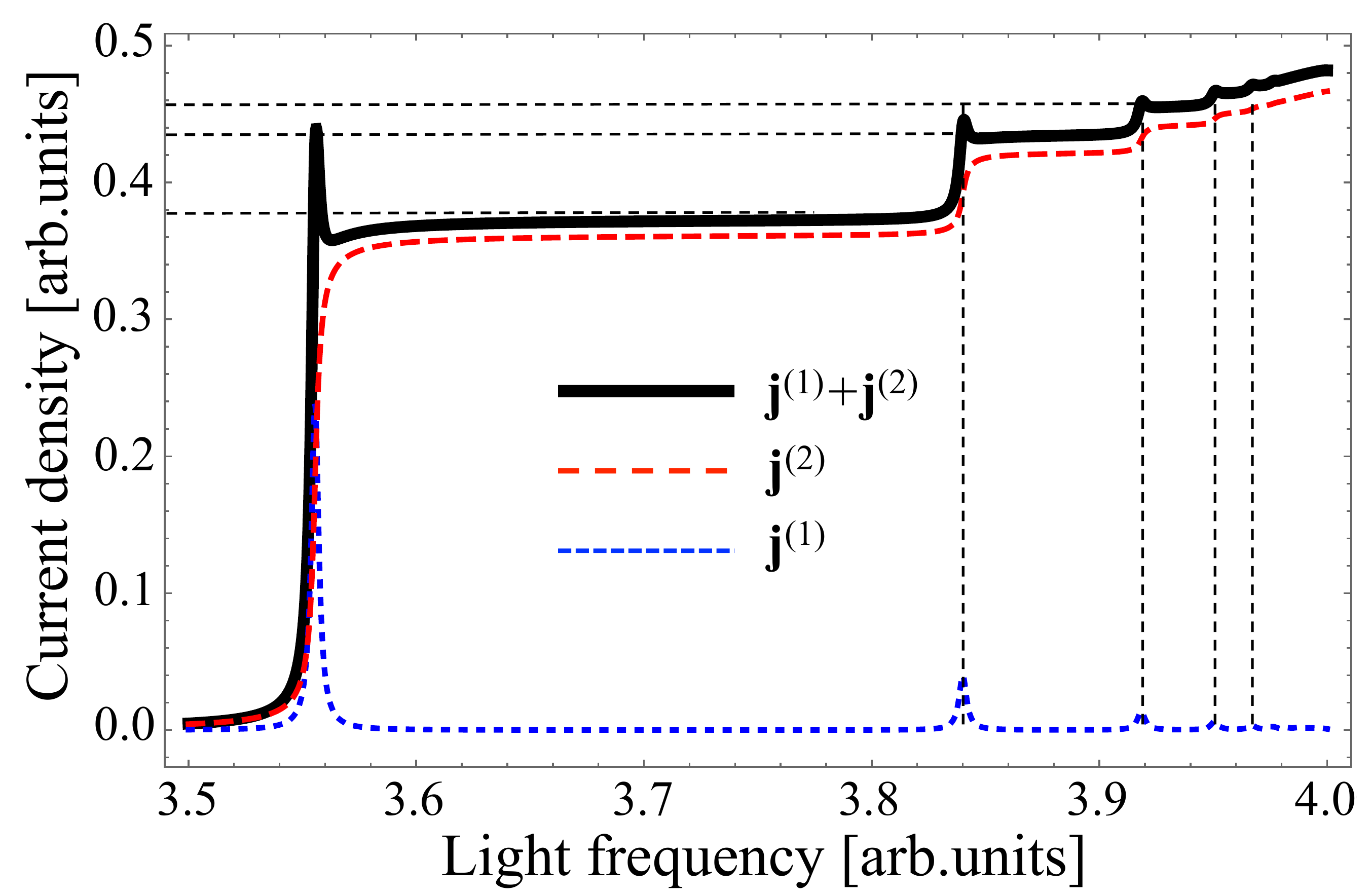}
\caption{Qualitative behaviour of the photoexcited part of exciton valley Hall current as a function of frequency of light (the red curve is slightly shifted down for better distinguishability). 
The calculations are performed within an analytically-tractable model of internal exciton spectrum~\cite{SM}. 
Red and blue curves correspond to the processes described by red and blue arrows in Fig.~\ref{Fig3}(a).}
\label{Fig4}
\end{figure}
As it is shown in Ref.~\onlinecite{ExcitonBerry2}, the Berry curvature influences these states resulting in their energy splitting of the order of tens of meV.
Moreover, the splitting has opposite sign in different valleys, see Fig.~\ref{Fig3}(b).

If the exciton gas is exposed to a circularly polarized electromagnetic field, the intra-excitonic transitions occur in one valley due to the valley-dependent optical selection rules, simultaneously exciting the state $m=+1$ or $m=-1$ in the same valley.
Thus, the photoexcited valley Hall current occurs in the system resulting in the exciton accumulation at the sample edge across $\textbf{F}_0$, see Fig.~\ref{Fig4}.
The latter can be detected at the sample boundary by measuring the polarization of luminescence.
Due to the step-like behaviour of the exciton Hall current, the intensity of the luminescence should inherit analogous step-like behaviour since its intensity is proportional to the density of excitons at the sample edge, determined by the value of the Hall current.

The last issue we want to address is the applicability of the Bogoliubov model.
Even though this model implies a weakly-interacting system, in the long-wavelength limit it works well even in the case of a strongly interacting Bose gas~\cite{Belyaev}. 
This is due to the low-energy excitations under strong interaction between the Bose particles also acquire linear (sound-like) dispersion (with some effective renormalized parameters).

\textbf{\textit{Conclusions.}} 
We have reported on the QAVHE of bosons.
We considered indirect exciton transport in two parallel layers of noncentrosymmetric two-dimensional materials and shown that in the presence of the Bose-Einstein condensate, there emerges a quantized valley Hall current of excitons due to the anomalous group velocity of exciton center-of-mass motion under the action of an external electromagnetic field.
This current experiences steplike behavior as a function of electromagnetic field frequency, mimicking the quantization of the internal exciton motion without external magnetic field.


\textit{{Acknowledgments.}} 
We acknowledge the financial support by the Russian Science Foundation (Project No.~17-12-01039) and the Institute for Basic Science in Korea (Project No.~IBS-R024-D1).






\end{document}